\documentstyle[preprint,aps]{revtex}
\begin{document}
\input{epsf}
\draft
\newfont{\form}{cmss10}
\newcommand{\e}{\varepsilon}
\renewcommand{\b}{\beta}
\newcommand{\unity}{1\kern-.65mm \mbox{\form l}}
\newcommand{\D}{D \raise0.5mm\hbox{\kern-2.0mm /}}
\newcommand{\A}{A \raise0.5mm\hbox{\kern-1.8mm /}}
\def\pmb#1{\leavevmode\setbox0=\hbox{$#1$}\kern-.025em\copy0\kern-\wd0
\kern-.05em\copy0\kern-\wd0\kern-.025em\raise.0433em\box0}

\def\D{\hbox{\hbox{${D}$}}\kern-1.9mm{\hbox{${/}$}}}
\def\kbar{\hbox{$k$}\kern-0.2truecm\hbox{$/$}}
\def\nbar{\hbox{$n$}\kern-0.23truecm\hbox{$/$}}
\def\pbar{\hbox{$p$}\kern-0.18truecm\hbox{$/$}}
\def\nhbar{\hbox{$\hat n$}\kern-0.23truecm\hbox{$/$}}
\newcommand{\dif}{\hspace{-1mm}{\rm d}}
\newcommand{\dil}[1]{{\rm Li}_2\left(#1\right)}
\newcommand{\diff}{{\rm d}}
\newcommand{\be}{\begin{equation}}
\newcommand{\ee}{\end{equation}}
\newcommand{\bea}{\begin{eqnarray*}}
\newcommand{\bean}{\begin{eqnarray}}
\newcommand{\eea}{\end{eqnarray*}}
\newcommand{\eean}{\end{eqnarray}}
\newcommand{\beal}{\begin{large} \begin{eqnarray*}}
\newcommand{\eeal}{\end{eqnarray*} \end{large}}
\newcommand{\beanl}{\begin{large} \begin{equation} \begin{array}{ll}}
\newcommand{\eeanl}{\end{array} \end{equation} \end{large}}
\newcommand{\bel}{\begin{large} \begin{equation}}
\newcommand{\eel}{\end{equation} \end{large}}
\newcommand{\nn}{\nonumber}
\newcommand{\z}{{\sl z}}
\newcommand{\x}{{\sl x}}
\newcommand{\y}{{\sl y}}
\newcommand{\spos}{\sum_{n=1}^{+\infty}}
\newcommand{\snneg}{\sum_{n=0}^{+\infty}}

\newcommand{\sneg}{\sum_{n=-\infty}^{-1}}
\newcommand{\som}{\sum_{n\neq 0}}
\newcommand{\somma}{\sum_{n=-\infty}^{+\infty}}

\newcommand{\nor}{\frac{1}{\sqrt{2L}}}
\newcommand{\kap}{n\frac{\pi}{L}}
\newcommand{\kapp}{|n|\frac{\pi}{L}}
\newcommand{\bn}{b_{n}}
\newcommand{\bnc}{b_{n}^{+}}
\newcommand{\bm}{b_{m}}
\newcommand{\bmc}{b_{m}^{+}}
\newcommand{\dn}{d_{n}}
\newcommand{\dnc}{d_{n}^{+}}
\newcommand{\dm}{d_{m}}
\newcommand{\dmc}{d_{m}^{+}}
\newcommand{\cn}{C_{n}}
\newcommand{\cnc}{C_{n}^{+}}
\newcommand{\cm}{C_{m}}
\newcommand{\cmc}{C_{m}^{+}}
\newcommand{\salfa}{\sigma_{\alpha}(t,\x)}
\newcommand{\salfac}{\sigma_{\alpha}^{+}(t,\x)}

\title{The Schwinger Model in Light-Cone Gauge}
\author{A. Bassetto ($^*$)}
\address{CERN, Theory Division, CH-1211 Geneva 23, Switzerland\\
INFN, Sezione di Padova, Padua, Italy}
\author{G. Nardelli}
\address{Dipartimento di Fisica, Universit\`a di Trento,
38050 Povo (Trento), Italy \\ INFN, Gruppo Collegato di Trento, Italy}
\author{E. Vianello}
\address{Dipartimento di Fisica ``G.Galilei", Via Marzolo 8 -- 35131
Padua, Italy}
\maketitle
\begin{abstract}
The Schwinger model, defined in the
space interval $-L \le x \le L$, with (anti)periodic boundary
conditions, is canonically quantized in the light-cone gauge $A_-=0$
by means of equal-time (anti)commutation
relations.  The transformation diagonalizing the complete Hamiltonian
is explicitly constructed, thereby giving spectrum, chiral
anomaly and condensate. The structures of Hilbert spaces related both
to free and to interacting Hamiltonians are completely exhibited.
Besides the usual massive field, two chiral massless fields are
present, which can be consistently expunged from the physical space by means
of a subsidiary condition of a Gupta--Bleuler type. The chiral
condensate does provide the correct non-vanishing value in the
decompactification limit $L \to \infty$.
\end{abstract}
\noindent
{\it PACS}: 11.10 Kk, 11.15-q. {\it Keywords}: Schwinger model, light-cone 
gauge.

\noindent
($^*$) On leave of absence from
Dipartimento di Fisica ``G.Galilei", Via Marzolo 8 -- 35131
Padova, Italy.
\vfill\eject

\narrowtext

\section{Introduction}
\noindent

The Schwinger model \cite{schwinger} is a celebrated theoretical
laboratory, possessing many non-trivial properties \cite{lowenstein}.
The literature is so copious
that we refrain from providing further references.

The reduction to two dimensions entails tremendous simplifications in quantum
electrodynamics and, as a consequence, the Schwinger model turns out to be the
only local gauge theory with non-trivial coupling,  where exact operatorial
solutions can be found. In addition, it is believed that  some interesting
features that this model exhibits persist in more realistic field theories.

As an example, all the problems related to the spontaneous breakdown of the
chiral symmetry, and the corresponding chiral anomaly, are very similar to the
ones in four-dimensional $QCD$ ($QCD_4$), with the obvious advantage that,
in this
simplified two-dimensional world, these phenomena  appear in a much more
transparent way. Moreover,
 the  infinite degeneracy of the vacuum state due
to chiral symmetry is analogous to that found in
$QCD_4$
so that one is
naturally led to introduce $\theta$-vacua.

Nevertheless, it should be emphasized that, although similar, the intimate
reasons of the appearance of these phenomena are deeply different in the two
theories.  It is common belief that the infrared behaviour of $QCD_4$ is
at the
root of the confining force between quarks and antiquarks. In the Schwinger
model, such a confining force is a ``free bonus'' of the theory:
electrodynamics
naturally confines in $1+1$ dimensions, due to the linear rise of the  Coulomb
potential. Consequently, the Schwinger model is  a formidable laboratory
to study the effects of confinement, but certainly not the causes.

The  aim  of this paper is to develop a rigorous operatorial treatment of the
 Schwinger model, when considered in the space interval $-L \le x \le L$, with
periodic boundary conditions for the vector field and antiperiodic ones
for fermions.
The compactification of the spatial dimension will naturally provide an
infrared regulator of the theory, without spoiling gauge invariance.

We choose to quantize the model  in the ``strong'' light-cone gauge (LCG)
$A_-=0$,  by imposing
an {\it equal-time} canonical algebra \cite{bas4},\cite{bas1}.
 To our knowledge, such a choice has never been considered so far.
There are several reasons that make the LCG particularly
interesting.
First of all, contrary to a common prejudice, this gauge is not
manifestly unitary when the theory is quantized on space-like surfaces. 
In addition, the direction in which the theory is compactified has to be 
carefully chosen. We recall indeed 
that axial gauges cannot be defined on compact
manifolds (i.e. compactifying both space and time) without introducing
singularities in the vector potentials, due to Singer's theorem;
nevertheless, partial compactifications are possible, provided they occur in a
direction different from the one specifying the gauge choice: in the present
case, to compactify the range of $x^-$ would be inconsistent.
On the other hand, equal-${\it time}$ quantization suggests
to regularize the theory with respect to
infrared singularities by compactifying the space
direction.

In the Schwinger model, a manifestly unitary formulation can instead be obtained
in the Coulomb gauge. Thus, as far as the content of physical degrees of 
freedom 
is concerned, LCG can be seen as an intermediate choice between Coulomb and
Lorentz gauges, but with the great advange that the Faddeev--Popov
sector decouples in the non-Abelian generalization of the model, making 
the transition to the non-Abelian case  
smoother.

When the theory in LCG is quantized by equal-time commutation relations, 
Gauss' law does not
hold strongly; rather, the Gauss operator obeys a free-field
equation  and entails the presence in the Fock space of unphysical
degrees of freedom. The vanishing of Gauss' operator has to be
imposed weakly, by a mechanism that is reminiscent
of the Gupta--Bleuler quantization scheme for
electrodynamics in Feynman gauge, but with the great advantage that it can be
naturally extended to the non-Abelian case.

In the present case,  thanks to compactification of the spatial dimension,
we have an even richer structure of the Hilbert space, as topological
excitations
(zero modes)  come into the game. 

The plan of the paper is as follows.

In Sect. 2 
we review the canonical quantization in LCG; we define the space of
states for the free theory as the direct product of a 
Fock space ${\cal F}_A$ containing the
frequency modes, obtained through repeated action of creation operators on
a Fock
vacuum, times a ``quantum-mechanical'' ${\cal L}^2$ space of the
zero-mode sector. Then we  introduce the interaction, namely we
consider the Schwinger
model in
LCG and provide the relevant equal-time commutation relations.

Section 3 is devoted to quantization at $t=0$; products of quantum operators
are  defined and regularized by means of gauge-invariant  point
splitting techniques. In particular, the expressions for the currents and
for the
kinetic terms occurring in the Hamiltonian are carefully discussed.

Section 4 deals with the diagonalization of the complete Hamiltonian.
The  unitary transformation, which diagonalizes the Hamiltonian,
relating free and
interacting fields, is explicitly exhibited. 
This is made possible by the partially compact
nature of the manifold (cylinder). The free-fields
sector contains, besides the ``physical'' massive scalar field with mass
$e/\sqrt\pi$, two additional ``unphysical''  chiral massless fields, with the
same chirality, but opposite signs in the commutators.
Consequently, the Fock
space will have an indefinite metric.

In Sect. 5 we discuss the temporal evolution of the fields. All the operatorial
solutions are explicitly exhibited and we check that 
they satisfy the correct Euler--Lagrange 
equations of motion. In the same section, it is also shown that 
Gauss' law is implemented by defining a physical Hilbert space in
which the
total charge vanishes and by selecting a suitable zero-norm combination of
ghost-like and ``longitudinal'' degrees of freedom, in the same way as 
the Gupta--Bleuler formalism in the Feynman gauge.

The structure of the ``physical'' Hilbert space is  discussed in
Sect. 6. We show that the same unitary operator diagonalizing the
Hamiltonian performs the mapping between free and interacting Fock spaces.
A set of degenerate vacua is obtained,
all of them being related by a residual gauge transformation. Then, the
requirement of  gauge invariance naturally leads to the definition of
$\theta$-vacua. Moreover a positive
definite inner product is derived, and it is shown that the
specific ghost-particle combinations 
forced by the (weakly) vanishing of Gauss'
operator always have vanishing norm. The physical Hilbert space is
eventually defined as
a quotient space.
Finally, chiral anomaly and chiral condensate are  discussed. As a final
check it is 
shown that the chiral condensate reproduces the correct non-vanishing
result of the continuum in the
decompactification limit $L\to \infty$.

\section{The Schwinger Model}

\subsection{The free-fermion model}

We start by 
recalling the treatment of free massless fermions in the space interval
$[-L,L]$. We assume antiperiodic boundary conditions $\psi (t,-L)=
-\psi(t,L)$. The classical Lagrangian is given by
\[ {\cal L }= i\bar{\psi}(x)\partial_{\mu}\gamma^{\mu}\psi(x). \]
Here $\psi$ is a two-component field:
 \[\psi=  
 \left(\begin{array}{cc}
\psi_{1}\\
\psi_{2}
\end{array}\right)\;, \] \\
$\gamma^{0}$ and $\gamma^{1}$ are $2\times 2$ matrices such that
the algebra
\[ \{\gamma^{\mu},\gamma^{\nu} \}=2g^{\mu \nu} \; \]
is fulfilled.

The action is invariant under global phase and chiral trasformations
leading to the well-known classical expression for the conserved vector
and axial vector currents.

When performing the canonical quantization
the classical variable $\psi(x)_{\alpha}$ and its conjugate momentum
\[ \pi_{\alpha}(x) \equiv 
\frac{\partial{\cal L}}{\partial(\:\partial_{0}\psi_{\alpha}(x)\;)}=
i\psi^{+}_{\alpha}(x) \]
become operators, obeying the canonical anticommutation relations
\bean
\{\psi_{\alpha}(t,\x),\psi_{\beta}^{+}(t,\y) \}
=\delta_{\alpha\beta} e^{{i\pi \over 2L}(\x-\y)}\delta(\x-\y)\ ,
\;\;\;\;\;\;\alpha,\beta=1,2 \ ,\nonumber 
\\ \label{anticomm} 
\{\psi_{\alpha}(t,\x),\psi_{\beta}(t,\y)\}=
\{\psi_{\alpha}^{+}(t,\x),\psi^{+}_{\beta}(t,\y)\}=0, 
\eean
where $\delta (\x-\y)$ denotes the periodic $\delta$-distribution.

Using the explicit representation for the gamma matrices
\( \gamma^{\mu}\; \;,\; \; \mu=0,1 \; : \) \\
\[\gamma^{0}=
\left(\begin{array}{cc}
0 & 1\\1 & 0
\end{array}\right)\; \;, \qquad \; \;
\gamma^{1}=
\left(\begin{array}{cc}
0 & 1\\-1 & 0
\end{array}\right)\ \ , \]
we get the equations of motion
\[
\partial_{-}\psi_{1}(x)=0 \ ,
\qquad \partial_{+}\psi_{2}(x)=0 \ ,
\]
where 
\bea
&&\x^{-}=\frac{t-\x}{\sqrt{2}} \; \;\;\;\;\; ,
\; \;\;\;\;\; \x^{+}=\frac{t+\x}{\sqrt{2}} \ ,\\ 
&&\partial_{-}=\frac{\partial}{\partial \x^{-}}=
\frac{\partial_{0}-\partial_{1}}{\sqrt{2}}\;\;\;\; \; \;,\; 
\;\;\;\;\;
\partial_{+}=\frac{\partial}{\partial \x^{+}}=
\frac{\partial_{0}+\partial_{1}}{\sqrt{2}}\;\;\; . 
\eea 

The most general solution fulfilling the condition
\be  \psi(t,\x)=-\psi(t,\x+2L) \label{condizione} 
\ee
is
\begin{eqnarray}
\psi_{1}(t,\x)&=&\frac{1}{\sqrt{2L}}
\snneg\ (\bn\,e^{-i(n+1/2)\frac{\pi}{L}(t+\x)}  
+\dnc\, e^{i(n+1/2)\frac{\pi}{L}(t+\x)}) \ , \nn \\
\label{ferm} \\
\psi_{2}(t,\x)&=&
\frac{1}{\sqrt{2L}}\sneg\ (\bn \, e^{i(n+1/2)\frac{\pi}{L}(t-\x)}
+\dnc\,e^{-i(n+1/2)\frac{\pi}{L}(t-\x)}) \ , \nn
\end{eqnarray}
(``left'' and ``right'' movers).

The canonical anticommutation relations induce the algebra
\be \begin{array}{cc}
&\{\bn ,\bmc \}=\delta_{m,n}\; , \;\;\;\;\; 
\{\dn\ ,\dmc \}=\delta_{m,n}\; , \;\;\;\;\; 
\{\bn\ ,\dmc \}=0 \ ,\\ 
&\{\bn\  ,\bm \}=\{\dn\ ,\dm \}=\{\bn\ ,\dm \}=0 \ .
\end{array}\label{fock} \ee 

The classical energy--momentum tensor
\be
\Theta^{\mu \nu }=
\frac{\partial\cal{L}}{\partial(\partial_{\mu}\psi_{\alpha})}
\partial^{\nu}\psi_{\alpha}-{\cal L}g^{\mu \nu}
\ee
leads to the corresponding conserved quantities
\be \label{qcons}
P^{0}\equiv H=\int_{-L}^{L}d\x\, \Theta^{00}(x) \;\;\;\;\;\;\;\; 
; \;\;\;\;\;\;\;\;
P^{1}=\int_{-L}^{L}d\x\, \Theta^{01}(x).
\ee

We get a quantum description by defining the products as normal-ordered
and introducing creation and annihilation operators:
\bean
 H=-\int_{-L}^{L}d\x\, 
:i\bar{\psi}(x)\gamma^{1}\partial_{1}\psi(x):\;= 
\somma k_{0}[n+1/2]\,(\bnc\ \bn\ +\dnc\ \dn),
\label{ham1}  \\
P_{1}=\int_{-L}^{L}d\x\, :i\psi^{+}(x)\partial_{1}\psi(x):
=\somma k_{1}[n+1/2]\,  (\bnc\ \bn\ +\dnc\  \dn) \label{p1} \ ,
\eean 
where
\be
k_{0}[n+1/2]= \left|n+\frac{1}{2}\right|\frac{\pi}{L} \quad ,\quad
k_{1}[n+1/2]= \left(n+\frac{1}{2}\right)\frac{\pi}{L} \ .
\ee 
Electric charge $Q$ and chiral charge
$Q_{5}$ are given by the following operators: 
\[ Q=e\int_{-L}^{L}d\x:\psi^{+}(x)\psi(x): \ , \]
\[ Q_{5}=-e\int_{-L}^{L}d\x:\psi^{+}(x)\gamma^{5}\psi(x): \ . \]
With
\[ \gamma^{5}=\gamma^{0}\gamma^{1}=
\left(\begin{array}{cc}
-1 & 0\\0 & 1
\end{array}\right)\;, \]
$Q$ and $Q_{5}$ become
\be \label{q}
Q=e\somma\ (\bnc\ \bn -\dnc\ \dn\ ) \ ,
\ee
\be \label{qcinque}
Q_{5}=e\snneg\ (\bnc\ \bn -\dnc\ \dn) -e\,\sneg\ (\bnc\ \bn -\dnc\ \dn) 
\;.
\ee 
We are now ready to introduce the bosonization.
From the fermionic currents
\bea
j^{-}=\frac{j^{0}-j^{1}}{\sqrt{2}}=e\sqrt{2}:\psi^{+}_{1}\psi_{1}: \ , \\ 
j^{+}=\frac{j^{0}+j^{1}}{\sqrt{2}}=e\sqrt{2}:\psi^{+}_{2}\psi_{2}: \ ,
\eea 
we get
\bean
&&j^{-}=\frac{e}{\sqrt{2}L}\sum_{p=1}^{\infty}
\Bigl( C_{p}e^{-i\frac{\pi}{L}p(t+\x)}
+C^{+}_{p}e^{i\frac{\pi}{L}p(t+x)}\Bigr)
+\frac{e}{\sqrt{2}L}\sum_{n=0}^{\infty}\Bigl(\bnc \bn -\dnc \dn \Bigr) \ , 
\\
&&j^{+}=\frac{e}{\sqrt{2}L}\sum_{p=-\infty}^{-1}\Bigl( C_{p}e^{
+i\frac{\pi}{L}p(t-\x)}
+C^{+}_{p}e^{-i\frac{\pi}{L}p(t-x)}\Bigr)
+\frac{e}{\sqrt{2}L}\sneg\ \Bigl(\bnc \bn -\dnc \dn \Bigr)\ , 
\eean
where
\bean
&&\cn =\sum_{m=0}^{n-1}\dm\ b_{n-m-1}+\sum_{m=0}^{\infty}(\bmc\ b_{m+n}
-\dmc\ d_{m+n})\;\;\;\; , \;\;{\rm for}\;\; n>0 \ , \nonumber \\
\label{bosoni} \\
&&\cn =\sum_{m=n}^{-1}\dm\ b_{n-m-1}
+\sum_{m=-\infty}^{-1}(\bmc\ b_{m+n}-\dmc\ d_{m+n})\;\; , \;\;\;\;{\rm for}\; n<0\ .
\nonumber
\eean 
One can easily verify that the fusion operators
$\cn\ $ obey the following {\it bosonic} commutation rules
\bean 
&&[\cn\ ,\cmc\ ]=|n|\delta_{m,n} \ ,
 \nn \\
&&[\cn\ ,\cm ]=[\cnc\ ,\cmc ]=0\ . \label{commbos}
\eean 
We notice that the operators $Q$ , $Q_{5}$ , $\cn$, act on the fermionic 
Fock space 
$\cal{F}$ obtained starting from a vacuum state $|0\rangle$ annihilated by
the operators $b_n$ and $d_n$, and that
\be
[\cn , Q]=[\cn , Q_{5}]=[Q , Q_{5}]=0.
\ee 
Let us now consider a peculiar set of states
$|M,N\rangle$ \cite{nakawaki} . 
The state $|M,N\rangle$ represents $M$ particles ($M<0$) or 
antiparticles
($M>0$) of the {\it left} type in the lowest $M$ energetic levels and
$N$ particles ($N<0$) or antiparticles ($N>0$) of the {\it right}
type in the lowest $N$ levels; namely
\bean 
&&|M,N\rangle=b_{M-1}^{+}\cdots
b_{0}^{+}b^{+}_{-N}\cdots b^{+}_{-1}|0\rangle \quad , \quad {\rm for}\quad M<0\; ,\;
 N<0\ , \nn \\ 
&&|M,N\rangle =d_{M-1}^{+}\cdots
d_{0}^{+}d^{+}_{-N}\cdots d^{+}_{-1}|0\rangle \quad  , 
\quad {\rm for} \quad M>0\; ,\;
 N>0\ ,  \\ \label{vuoti} 
&&|M,N\rangle =d_{M-1}^{+}\cdots
d_{0}^{+}b^{+}_{-N}\cdots b^{+}_{-1}|0\rangle \quad ,  
\quad {\rm for} \quad M>0\; , \;
 N<0\ , \nn \\ 
&&|M,N\rangle =b_{M-1}^{+}\cdots
b_{0}^{+}d^{+}_{-N}\cdots d^{+}_{-1}|0\rangle \quad ,  
\quad {\rm for} \quad M<0 \;,\; 
N>0\ . \nn 
\eean
Since
\bean
&&Q|M,N\rangle =-e(M+N)|M,N\rangle ,\\
&&Q_{5}|M,N\rangle =-e(M-N)|M,N\rangle ,
\eean
any given state $|M,N\rangle $ can be identified by means of the
eigenvalues of $Q$ and $Q_{5}$ . \\
It can be verified that for any state $|M,N\rangle $ 
\[ C_{n}|M,N\rangle =0 \; .\]
Moreover, as the operators
$C_{n}^{+}$ commute with the charges, the eigenvalues of $Q$ and $Q_{5}$
are not modified by applying 
$C_{n}^{+}$ on the states $|M,N\rangle $.

In ref. \cite{uhlenbrock} it has been shown that the fermionic Fock space 
${\cal {F}}$ can be decomposed as an infinite direct sum of irreducible Fock
representations of the algebra
(\ref{commbos}), which coincide with the eigenspaces of
$Q$ and $Q_{5}$. 

The Hamiltonian and the momentum (eqs. (\ref{ham1}) and (\ref{p1}))
can be expressed in terms of the charges and of the fusion operators through:
\bean 
&&H=\frac{\pi}{4Le^{2}}\left(Q^{2}+Q_{5}^{2}\right)+\frac{\pi}{L}\som\ 
\cnc \cn \ ,
\label{ham2}  \\
&& P_1=\frac{\pi}{2Le^{2}}Q\ Q_{5}+{\pi \over L} 
\som\ \epsilon(n)\  \cnc \cn \ . \label{p2}
\eean 

In order to express the fermionic operators in terms of the fusion operators,
it is useful to introduce the quantities
\bean 
\varphi_{1}^{(+)}(t,\x)&=&-\spos\ \frac{1}{n}{\cn}e^{-i\kap(\x+t)} \ , 
\nonumber \\
\varphi_{1}^{(-)}(t,\x)&=&\spos\ \frac{1}{n}{\cnc}e^{i\kap(\x+t)} \ , 
\\  
\varphi_{2}^{(+)}(t,\x)&=&\sneg\ \frac{1}{n}{\cn}e^{i\kap(t-\x)} \ ,
\nonumber \\
\varphi_{2}^{(-)}(t,\x)&=&-
\sneg\ \frac{1}{n}{\cnc}e^{-i\kap(t-\x)}\ .\nonumber
\eean
Then we can define the operators
\be 
\sigma_{\alpha}(t,\x)=\sqrt{2L}e^{\varphi_{\alpha}^{(-)}(t,\x)}
\psi_{\alpha}(t,\x)e^{\varphi_{\alpha}^{(+)}(t,\x)},
\ee 
which carry the fermionic charge.
One can easily derive the relations\cite{nakawaki} : 
\be
\label{grazia1}
\sigma_{\alpha}^{+}(t,\x)\sigma_{\alpha}(t,\x)
=\sigma_{\alpha}(t,\x)\sigma_{\alpha}^{+}(t,\x)=1 \ ,
\ee
\be 
\label{grazia2}
\{\sigma_{1}(x),\sigma_{2}(y)\}=
\{\sigma_{1}(x),\sigma^{+}_{2}(y)\}=0 \ ,
\ee
\be \label{24}
[Q,\salfa]=-e\salfa \;\;\;\; , \;\;\;\;[Q_{5},\salfa]=(-1)^{\alpha}e\salfa \ ,
\ee 
\be \label{25}
[\cn,\salfa]=[\cnc,\salfa]=0 \ .
\ee 
The quantities 
$\sigma_{\alpha}\equiv \sigma_{\alpha}(0,0)$
are usually called {\it spurions}.

Using eqs.
(\ref{ham2}), (\ref{p2}), (\ref{24}) and
(\ref{25}), it can be shown that 
\[ \salfa
=e^{\Lambda_{\alpha}(t,\x)}\sigma_{\alpha}e^{\Lambda_{\alpha}(t,\x)}\ ,\]
where
\[ \Lambda_{\alpha}(t,\x)\equiv -\frac{i\pi}{4Le}[Q-(-1)^{\alpha}
Q_{5}][t-(-1)^{\alpha}\x] \ . \]
The following relation can be verified:
\be
|M,N\rangle =\sigma_{1}^{M}\sigma_{2}^{N} |0\rangle \ ,
\ee
where $\sigma_{\alpha}^{-1} = \sigma_{\alpha}^+$.
It is useful to rescale the fusion operators
\be
C_{n}=i\sqrt{|n|}\gamma_{n} \quad ,\quad
C_{n}^{+}=-i\sqrt{|n|}\gamma^{+}_{n} \ ,
\ee
so that
\be \label{commgamma}
[\gamma_{n},\gamma^{+}_{m}]=\delta_{m,n}\quad , \quad
[\gamma_{n},\gamma_{m}]=[\gamma_{n}^{+},\gamma_{m}^{+}]=0  \;.
\ee
Defining the bosonic field
\bean
\varphi(x)&=&\frac{i}{2\sqrt{\pi}}\left(\varphi_{1}^{(+)}+
\varphi_{1}^{(-)}-\varphi_{2}^{(+)}-\varphi_{2}^{(-)}\right)= \nn \\ 
&=&\nor \som \frac{1}{\sqrt{2k_{0}}}
\Bigl(\gamma_{n} e^{-ik\cdot x}+\gamma_{n}^{+}e^{ik\cdot x} \Bigr) \ , 
\eean 
the fermionic currents can be rewritten as
\bean
&&j^{-}=\frac{Q+Q_{5}}{2\sqrt{2}L}-\frac{e}{\sqrt{\pi}}\partial_{+}
\varphi \ ,  \\ 
&&j^{+}=\frac{Q-Q_{5}}{2\sqrt{2}L}-\frac{e}{\sqrt{\pi}}\partial_{-}
\varphi \ ,
\eean 
where $\varphi$ is a `quasi-scalar' field, as it has no zero mode.

\subsection{The free Maxwell field at $t=0$}

As we are considering the theory on the cylinder, we can assume that 
the algebra of interacting
fields at a given time ($t=0$) is isomorphic to the free-fields algebra.
\footnote{We recall that this assumption would not be allowed in
the continuum, owing to Haag's theorem.}

The free Maxwell field is described by the Lagrangian
\be 
\label{lagr1} 
{\cal L} =-\frac{1}{4}F_{\mu\nu}F^{\mu\nu}.
\ee 
In order to quantize it we have to choose a gauge; we choose the light-cone 
gauge by adding to the classical Lagrangian the following 
``gauge-fixing'' term 
\[ {\cal L}_{gf}=\lambda^{(0)} n_{\mu} A^{\mu} \; ,\]
where $n_{\mu}= {1\over{\sqrt{2}}}(1,1)$ and $\lambda^{(0)}$ 
is a Lagrange multiplier
enforcing the gauge condition $n_{\mu} A^{\mu}=0$.

We define
\[ A=A_{+}=\frac{A_{0}+A_{1}}{\sqrt{2}} \;\;\;,\;\;\;
F=F_{01} \;.\]

The Euler--Lagrange equations are :
\be
\partial_{\mu}F^{\mu \nu}+ n^{\nu}{\lambda^{(0)}}=0 \label{gauss+lambda1} ,
\ee 
which, in turn, imply
\be 
\partial_{-}\lambda^{(0)}=0.
\ee

Following the standard Dirac procedure, we can derive the following
independent {\it equal-time} commutation relation:
\be
[A(t,\x),F(t,\y)]=i\sqrt{2}\delta(\x-\y). 
\ee 

At variance with the Coulomb gauge choice, equal-time quantization
in the light-cone gauge entails the presence of unphysical degrees of freedom
\cite{bas1}. Strictly speaking, in order to recover Maxwell's equations,
one should impose the condition
$\lambda^{(0)}=0$. However, as $\lambda^{(0)}$ does not commute 
with $A$, 
to impose $\lambda^{(0)}=0$ in an operatorial sense would be inconsistent
with quantization. In analogy with the Gupta--Bleuler formalism, this
condition will be imposed weakly.

We start considering the quantization of the bosonic field at $t=0$.

The Fourier expansion of $A(0,\x)$ and $F(0,\x)$ is 
\bean
&&A(0,\x)=\sqrt{2}A_{1}(0,\x)=\frac{1}{\sqrt{L}}\som a_{n}
e^{in\frac{\pi}{L}\x}+\frac{{\bf a}_{0}}{\sqrt{L}} \ , \label{A}   \\
&&F(0,\x)=\frac{1}{\sqrt{2L}}\som b_{n}
e^{in\frac{\pi}{L}\x}+\frac{{\bf b}_{0}}{\sqrt{2L}} \ , \label{F} \\
&&a_{n}^{+}=a_{-n} \quad ,\quad b_{n}^{+}=b_{-n} \ . \nn 
\eean 
From the canonical commutation relations of $A$ and $F$ it follows that
\be \label{ab}
[a_{n},b_{m}]=i\delta_{n,-m} \ ,
\ee
\be \label{a0b0}
[{\bf a}_{0},{\bf b}_{0}]=i ,
\ee
all other commutators vanishing.

As far as zero modes are concerned, we represent
(\ref{a0b0}) in a ${\cal L}^{2}$-space by means of the relations:
\[
{\bf a}_{0}\psi(a_{0})=a_{0}\psi(a_{0}) \qquad ,\qquad
{\bf b}_{0}\psi(a_{0})=-i \frac{d}{da_{0}}\psi(a_{0})\qquad ,
\qquad \psi\in {\cal L}^{2} .
\]

For the frequency modes we shall consider a Fock space ${\cal F}_A$ 
representation in
terms of suitable creation and annihilation operators.
The free bosonic Hamiltonian is: \\
\[
H^{(0)}=\int_{-L}^{L} d\x \left(\frac{1}{2}{F}^{2}(0,\x)
-\frac{A(0,\x)}{\sqrt{2}}\partial_{1}F(0,\x) \right) \; .
\]
The operator ordering in this equation is irrelevant in obtaining the
commutators that follow.
From 
(\ref{A}) and (\ref{F}) we get
\[
H^{(0)}=\frac{1}{2}\som b_{n}b_{-n}-i\som k_{1}a_{-n}b_{n}
+\frac{1}{2}b_{0}^{2}
\]
and thereby
\bea
&&[H^{(0)},b_{n}]=k_{0}b_{n}\ , \qquad {\rm for} \quad  n>0 \ , \\ 
&&[H^{(0)},b_{n}]=-k_{0}b_{n}\ , \qquad {\rm for}  \quad n<0 \ ,
\eea
which suggest an interpretation of $b_n$ as an annihilation operator for $n<0$
and as a creation operator for $n>0$. The commutators for
$a_n$ follow from analogous considerations.

${\cal F}_{A}$ is now constructed by means of the following
creation and annihilation operators:
\bean
A_{n}=a_{n}\sqrt{\frac{k_{0}}{2}}+i\frac{b_{n}}{\sqrt{2k_{0}}}
\qquad , \qquad 
A_{n}^{+}=a_{-n}\sqrt{\frac{k_{0}}{2}}-i\frac{b_{-n}}{\sqrt{2k_{0}}}\ ,
\qquad  {\rm for} \quad n<0 \ ; 
&&  \\
A_{n}=a_{-n}\sqrt{\frac{k_{0}}{2}}-i\frac{b_{-n}}{\sqrt{2k_{0}}}
\qquad , \qquad 
A_{n}^{+}=a_{n}\sqrt{\frac{k_{0}}{2}}+i\frac{b_{n}}{\sqrt{2k_{0}}}\ ,
\qquad  {\rm for} \quad n>0 \ . \label{An}
\eean 
From (\ref{ab}) it follows that
\bean
 &[A_{n},A_{m}^{+}]=\delta_{nm}\ , \qquad &{\rm for}\quad n,m<0 \ ,  \label{+}
\\ 
 &[A_{n},A_{m}^{+}]=-\delta_{nm}\ , \qquad &{\rm for} \quad n,m>0 \ ,
\label{-}
\eean 
all other commutators vanishing.

We notice that, since $A_{n}$ is an annihilation operator and 
$A_{n}^{+}$ is a creation one,
the negative sign in
(\ref{-}) entails the presence of negative norm states in the Fock space 
${\cal F}_A$. On the other hand we know that redundant degrees of freedom
are present, as we have to impose the vanishing of $\lambda^{(0)}$, the 
superscript $(0)$ denoting free fields.
As a consequence we have to consistently define a physical Hilbert space.
Let us consider the time evolution of the Gauss operator
\[\lambda^{(0)}(t,\x)=-\sqrt{2}\partial_{1}F^{(0)}(t,\x)=
-\sqrt{2}e^{-iH_{0}t}\partial_{1}F(0,\x)e^{iH_{0}t}  \; .\]

We get
\bea
\lambda^{(0)}(t,\x)&=&\frac{1}{\sqrt{2L}}\Biggl\{\sneg k_{0}\sqrt{k_{0}}
(A_{n}-A_{-n})e^{-ik_{0}(t+\x)}\\ 
&& \qquad \qquad -\spos k_{0}\sqrt{k_{0}}(A_{n}^{+}-A_{-n}^{+})
e^{ik_{0}(t+\x)}\Biggr\}\ .
\eea 
The physical states are defined by
\[ {\lambda^{(0)}}^{(+)}|phys \rangle =0 \ ,\]
namely
\be \label{phys}
(A_{n}-A_{-n})|phys \rangle =0\ , \qquad \forall n < 0 \ ,
\ee
which is reminiscent of the Gupta--Bleuler condition in the Feynman gauge.

In the vector space of linear combinations of products of creation
operators acting on the vacuum, we introduce a positive-definite inner product
$\langle\cdot , \cdot \rangle $ \ 
such that \ $[A_{n},A^{*}_{m}]=\delta_{nm}$,
where $A_{n}^{*}$ is the adjoint of $A_{n}$ with respect to the inner
product $\langle\cdot , \cdot \rangle $. A Fock space can be constructed by
a Cauchy completion with respect to this inner product, which is 
related to the product
\ $(\cdot ,  \cdot) $ \ 
compatible with (\ref{-}), by:
\[ (\cdot ,  \cdot) 
=\langle\cdot ,\eta \; \cdot \rangle \ ,\]
where
\[\eta=(-1)^{N} \qquad , \qquad N=-\sum_{n<0}A_{-n}^{+}A_{-n} \ . \]
In turn the inner product \ $(\cdot ,  \cdot)$ \  
in the physical subspace defined by (\ref{phys}) turns out to be 
semi-positive-definite.
The physical Hilbert space is eventually obtained as the quotient space.

\subsection{The Schwinger model in the light-cone gauge}
The Schwinger model is defined by the following Lagrangian 
\be 
\label{lagr} 
{\cal L} =-\frac{1}{4}F_{\mu\nu}F^{\mu\nu}+
\frac{i}{2}(\bar{\psi}\partial_{\mu}\gamma^{\mu}\psi-
\partial_{\mu}\bar{\psi}\gamma^{\mu}\psi)-j^{\mu}A_{\mu} \;.
\ee 
In order to quantize it we have to choose a gauge; we choose the light-cone 
gauge by adding to the classical Lagrangian the following 
``gauge-fixing'' term 
\[ {\cal L}_{gf}=\lambda n_{\mu} A^{\mu} \; ,\]
where $n_{\mu}= {1\over{\sqrt{2}}}(1,1)$ and $\lambda$ is a Lagrange multiplier
enforcing the gauge condition $n_{\mu} A^{\mu}=0$.

The classical gauge-fixed Hamiltonian is
\be \label{H}
H=\int_{-L}^{L} d\x \left(\frac{1}{2}F^{2}(x)-\frac{A}{\sqrt{2}}
\partial_{1}F(x) -\frac{1}{2}\Bigl[i\bar{\psi}(x)
\gamma^{1}\partial_{1}\psi(x)-i\partial_{1}\bar{\psi}(x)\gamma^{1}\psi(x)
\Bigr] + A(x)j^{+}(x) \right) ,
\ee 
where, again,
\[ A=A_{+}=\frac{A_{0}+A_{1}}{\sqrt{2}} \;\;\;,\;\;\;
F=F_{01} \;.\]

The Euler--Lagrange equations are :
\be
\partial_{\mu}F^{\mu \nu}-j^{\nu}+ n^{\nu}{\lambda}=0 \label{gauss+lambda} ,
\ee 
which, in turn, imply
\be 
\partial_{-}\lambda=0.
\ee

Following the standard Dirac procedure, one can derive the following
independent {\it equal-time} (anti)commutation relations:
\bea
&&[A(t,\x),F(t,\y)]=i\sqrt{2}\delta(\x-\y), \\ 
&&\{\psi(t,\x),\psi^{+}(t,\y)\}=e^{{i\pi \over 2L}(x-y)}\delta(\x-\y),  
\eea 
all other (anti)commutators vanishing.

Again, as $\lambda$ does not commute with $A$ and $\psi$, 
the condition $\lambda=0$ will be imposed weakly.

\section{Regularization of composite quantum operators at equal time}
\subsection{Point splitting regularization}
Products of operators are plagued by ultraviolet singularites that
need to be regularized. In the following we shall use the standard 
point splitting technique. To preserve gauge invariance, the following bosonic
string 
\[ e^{-ie\int_{x}^{x+\varepsilon} dy^{\nu}A_{\nu}(y)} \]
will be inserted in the fermionic bilinears. The exponential involves 
products of operators at the same point, which in turn have to be
suitably regularized. This problem will be carefully discussed
when needed. 
For the time being, we {\it formally} define
\cite{johnson}
\bean
&&j^{\mu}(x)=\frac{e}{2}\lim_{\varepsilon \to 0}
 \,\bigg[ \bar{\psi}(x+\varepsilon)\gamma^{\mu}
e^{-ie\int_{x}^{x+\varepsilon} dy^{\nu}A_{\nu}(y)}\psi(x)+\nn 
\\ 
&+&\bar{\psi}(x)\gamma^{\mu}
e^{-ie\int^{x}_{x+\varepsilon} dy^{\nu}A_{\nu}(y)}
\psi(x+\varepsilon) \bigg]\ ,\qquad \varepsilon^{2}< 0 \label{corrente} 
\eean  
and 
\bean
&&\Big[i\bar{\psi}(x)\gamma^{\mu}\partial_{\mu}\psi(x)
-i\partial_{\mu}\bar{\psi}(x)\gamma^{\mu}\psi(x)\Big]_{R}= 
\lim_{\varepsilon \to 0} 
\bigg(i\bar{\psi}(x+\varepsilon)\gamma^{\mu}
e^{-ie\int_{x}^{x+\varepsilon} dy^{\nu}A_{\nu}(y)}\partial_{\mu}\psi(x)- 
\nn \\
&-&\ i\partial_{\mu}\bar{\psi}(x)\gamma^{\mu}
e^{-ie\int^{x}_{x+\varepsilon} dy^{\nu}A_{\nu}(y)}
\psi(x+\varepsilon)-v.e.v. \bigg)\ ,\qquad \varepsilon^{2}< 0\ , 
\label{terminecinetico}  
\eean 
where the vacuum considered here is the one of the fully interacting
theory.
The regularized quantum Dirac equation is 
\be
\label{checca}
i\gamma^{\mu}\partial_{\mu}\psi(x)
-e\gamma^{\mu}\Bigl[A_{\mu}(x)\psi(x)\Bigr]_{R}=0 \;,
\ee
where\cite{lowenstein} 
\[ \Bigl[A_{\mu}(x)\psi(x)\Bigr]_{R}
=\frac{1}{2}\lim_{\varepsilon \to 0}
\Bigl\{A_{\mu}(x+\varepsilon)\psi(x)
+\psi(x)A_{\mu}(x-\varepsilon)\Bigr\} \; ,\qquad \varepsilon^{2}< 0\ . \]

\subsection{Regularized current}

We are now in the position of giving an explicit expression for
the quantum fermionic currents.
We recall the definition
\[j^{-}(0,\x)=\frac{1}{2}\lim_{\epsilon\to 0}\,
\Big [\sqrt{2}e \psi^{+}_{1}(0,\x+\epsilon)
e^{-ie\int_{\x}^{\x+\epsilon}A_{1}(0,\y)d\y}\psi_{1}(0,\x)+
h.c. \Big] \ .\] 
As $A_{1}$ and $\psi_{1}$ commute at equal time, we can write
\[j^{-}(0,\x)=
\frac{1}{2}\lim_{\epsilon\to 0}\Big[j_{(0)}^{-}(0,\x;\epsilon) 
e^{-ie\int_{\x}^{\x+\epsilon}A_{1}(0,\y)d\y}+h.c. \Big] \ ,\] 
where
\be \label{3.16}
j_{(0)}^{-}(0,\x;\epsilon)\equiv 
e\sqrt{2}\psi_{1}^{+}(0,\x+\epsilon)\psi_{1}(0,\x) \ .
\ee
Using
\[ \psi_{1}(0,\x)=e^{-\varphi_{1}^{(-)}(0,\x)}e^{\Lambda_{1}(0,\x)}
\sigma_{1}e^{\Lambda_{1}(0,\x)}e^{-\varphi_{1}^{(+)}(0,\x)}\]
and
\[\Lambda_{1}(0,\x)=-\frac{i\pi}{4Le}(Q+Q_{5})\x \ ,\] 
it follows that
\[ \left[\varphi_{1}^{(-)}\right]^{+} =-\varphi_{1}^{(+)}\ , \qquad 
\Lambda_{1}^{+}=-\Lambda_{1} \ ,\]
and hence
\bea
j_{(0)}^{-}
={\textstyle\frac{e}{\sqrt{2}L}}\;e^{\varphi_{1}^{(-)}(0,\x+\epsilon)}
e^{-\Lambda_{1}(0,\x+\epsilon)}
\sigma_{1}^{+}e^{-\Lambda_{1}(0,\x+\epsilon)}
e^{\varphi_{1}^{(+)}(0,\x+\epsilon)} \cdot \\
\cdot e^{-\varphi_{1}^{(-)}(0,\x)}e^{\Lambda_{1}(0,\x)}
\sigma_{1}e^{\Lambda_{1}(0,\x)}e^{-\varphi_{1}^{(+)}(0,\x)} \ .
\eea 
By recalling eqs. (\ref{grazia1}) to (\ref{25}) 
and the 
relation
\be
e^{A}B=Be^{A}e^{c}\ , \qquad {\rm if}\quad [A,B]=cB\ , \quad {\rm with} \quad
c \quad {\rm being} \quad {\rm a}\quad c\ {\rm number} \ , 
\ee
we obtain
\bea
j_{(0)}^{-}
={\textstyle\frac{e}{\sqrt{2}L}}\;
e^{-\frac{i\pi}{2L}\epsilon}
e^{-2\Lambda_{1}(0,\epsilon)}
e^{\varphi_{1}^{(-)}(0,\x+\epsilon)}
e^{\varphi_{1}^{(+)}(0,\x+\epsilon)}
e^{-\varphi_{1}^{(-)}(0,\x)}
e^{-\varphi_{1}^{(+)}(0,\x)}.
\eea 
From the identity 
\be
e^{A}e^{B}=e^{B}e^{A}e^{[A,B]}\ , \qquad   {\rm if} \quad [A,B]=c\ {\rm 
number} \ ,
\ee
we can write
\[
e^{\varphi_{1}^{(+)}(0,\x+\epsilon)} 
e^{-\varphi_{1}^{(-)}(0,\x)}=
e^{-\varphi_{1}^{(-)}(0,\x)}e^{\varphi_{1}^{(+)}(0,\x+\epsilon)} 
e^{[\varphi_{1}^{+}(0,\x+\epsilon),-\varphi_{1}^{(-)}(0,\x)]} \ .
\] 
Using the decomposition for the quasi-scalar field $\varphi$, we get
\be \label{delta1}
[\varphi_{1}^{+}(0,\x+\epsilon),-\varphi_{1}^{(-)}(0,\x)]=
\spos \frac{1}{n}e^{-i\kap \epsilon}=-{\rm ln}\left(1-
e^{-i\frac{\pi}{L}\epsilon}\right) \ ,
\ee
and therefore
\be \label{j^-_0} 
j_{(0)}^{-}={\textstyle\frac{e}{\sqrt{2}L}}
\frac{e^{-\frac{i\pi}{2L}\epsilon}}{1-e^{-i\frac{\pi}{L}\epsilon}}
e^{-2\Lambda_{1}(0,\epsilon)}
e^{\bigl(\varphi_{1}^{(-)}(0,\x+\epsilon)-\varphi_{1}^{(-)}(0,\x)\bigr)}
e^{\bigl(\varphi_{1}^{(+)}(0,\x+\epsilon)-\varphi_{1}^{(+)}(0,\x)\bigr)} \ .
\nn
\ee
Neglecting terms that vanish in the limit \ $\epsilon\to 0$ \ 
we get
\[j_{(0)}^{-}(0,\x;\epsilon)=\frac{e}{i\pi\epsilon\sqrt{2}}+
\frac{Q+Q_{5}}{2\sqrt{2}L}+\frac{e}{i\pi\sqrt{2}}\Bigl(\partial_{1}
\varphi_{1}^{(-)}(0,\x)+\partial_{1}\varphi_{1}^{(+)}(0,\x)\Bigr). \]
Moreover
\[e^{-ie\int_{\x}^{\x+\epsilon}A_{1}(0,\y)d\y}
=1-ie\epsilon A_{1}(0,\x)+\ldots  \]
and therefore
\bea
j_{(0)}^{-}(0,\x;\epsilon) 
e^{-ie\int_{\x}^{\x+\epsilon}A_{1}(0,\y)d\y}&=&
\frac{e}{i\pi\epsilon\sqrt{2}}-\frac{e^{2}A_{1}(0,\x)}{\sqrt{2}\pi}+
\frac{Q+Q_{5}}{2\sqrt{2}L} + \\ \\ 
&&+\frac{e}{i\pi\sqrt{2}}\Bigl(\partial_{1}
\varphi_{1}^{(-)}(0,\x)+\partial_{1}\varphi_{1}^{(+)}(0,\x)\Bigr) \ .
\eea \\
From the definition of\ $j^{-}(0,\x)$ \ it follows that
\be
j^{-}(0,\x)=-\frac{e^{2}A_{1}(0,\x)}{\sqrt{2}\pi}+\frac{Q+Q_{5}}{2\sqrt{2}L}+
\frac{e}{\sqrt{2}L}\spos\left(i\sqrt{|n|}\gamma_{n}
 e^{-i\kap\x}-i\sqrt{|n|}\gamma_{n}^{+}
e^{i\kap\x}\right) \ .
\ee 
An analogous calculation gives, taking
\be
[\varphi_{2}^{(+)}(0,\x+\epsilon)\, , \, -\varphi_{2}^{(-)}(0,\x)]
=\spos\frac{1}{n} e^{i\kap\epsilon}=-ln\left(1-e^{i\frac{\pi}{L}\epsilon}
\right)
\ee
into account,
\bean
j^{+}(0,\x)&=&\lim_{\epsilon\to 0}\,
\Big [\sqrt{2}e \psi^{+}_{2}(0,\x+\epsilon)
e^{-ie\int_{\x}^{\x+\epsilon}A_{1}(0,\y)d\y}\psi_{2}(0,\x)+
h.c. \Big]= \nn \\ 
&=&\frac{e^{2}A_{1}(0,\x)}{\sqrt{2}\pi}+\frac{Q-Q_{5}}{2\sqrt{2}L}+
\frac{e}{\sqrt{2}L}\sneg\left(i\sqrt{|n|}\gamma_{n}e^{-i\kap\x}
-i\sqrt{|n|}\gamma_{n}^{+}e^{i\kap\x}\right)\ . 
\label{j+}
\eean 

\subsection{Regularized kinetic term}
The fermionic kinetic term entering the quantum Hamiltonian is
\bean
\lefteqn{H_{\psi}=-\frac{1}{2}\int_{-L}^{L} d\x \Bigl[i \bar{\psi}(0,\x)
\gamma^{1}\partial_{1}\psi(0,\x)-i\partial_{1}\bar{\psi}(0,\x)
\gamma^{1}\psi(0,\x) \Bigr]_{R}= } &&  \nn \\
&& \qquad \qquad =\frac{1}{2}\int_{-L}^{L} d\x
\Bigl[i\psi^{+}_{1}(0,\x)\partial_{1}\psi_{1}(0,\x)
-i\psi^{+}_{2}(0,\x)\partial_{1}\psi_{2}(0,\x)+h.c.\Bigr]_{R} \ , \nn
\eean 
where
\bea
\lefteqn{\Bigl[i\psi^{+}_{1}(0,\x)\partial_{1}\psi_{1}(0,\x)
-i\psi^{+}_{2}(0,\x)\partial_{1}\psi_{2}(0,\x)+h.c.\Bigr]_{R}=}&&  \\
&&=\lim_{\epsilon \to 0}
\biggl\{ i\psi^{+}_{1}(0,\x+\epsilon)e^{-ie\int_{\x}^{\x+\epsilon}
A_{1}(0,\y)d\y}
\partial_{1}\psi_{1}(0,\x)\\ 
&&
-i\psi^{+}_{2}(0,\x+\epsilon)e^{-ie\int_{\x}^{\x+\epsilon}A_{1}(0,\y)d\y}
\partial_{1}\psi_{2}(0,\x)
+h.c.-v.e.v. \biggr\}.
\eea  \\
Setting \ $\z=\x+\epsilon $, we can write
\bea
\lefteqn{\Bigl[i\psi^{+}_{1}(0,\x)\partial_{1}\psi_{1}(0,\x)
-i\psi^{+}_{2}(0,\x)\partial_{1}\psi_{2}(0,\x)+h.c.\Bigr]_{R}=}&&  \\
&&=\lim_{\z \to \x}
\left\{e^{-ie\int_{\x}^{\z}A_{1}(0,\y)d\y}
\frac{\partial}{\partial \x}\left[i\psi^{+}_{1}(0,\z)\psi_{1}(0,\x)-
i\psi^{+}_{2}(0,\z)\psi_{2}(0,\x)\right]
+h.c.-v.e.v. \right\} \ .
\eea 
Equation (\ref{3.16}) entails
\[ \psi^{+}_{1}(0,\z)\psi_{1}(0,\x)=
\frac{1}{e\sqrt{2}}j^{-}_{(0)}(\x,\z-\x) \] 
and, using (\ref{j^-_0}), one gets
\( 2L\frac{\partial}{\partial \x}
\left[\psi^{+}_{1}(0,\z)\psi_{1}(0,\x)\right]= \)
\[={\textstyle \frac{\partial}{\partial \x}}
\Bigg[\,e^{-\frac{i\pi}{2L}(\z-\x)}\,
\frac{1}{1-e^{-\frac{i\pi}{L}(\z-\x)}}\,
e^{-2\Lambda_{1}(0,\z-\x)} 
\,:e^{\left(\varphi_{1}(0,\z)-\varphi_{1}(0,\x)\right)}:\,\Bigg] \ ,\]
where 
\[\varphi_{1}=\varphi_{1}^{(+)}+\varphi_{1}^{(-)} \]
and
\[:e^{\left(\varphi_{1}(0,\z)-\varphi_{1}(0,\x)\right)}:=
e^{\left(\varphi_{1}^{(-)}(0,\z)-\varphi_{1}^{(-)}(0,\x)\right)}
e^{\left(\varphi_{1}^{(+)}(0,\z)-\varphi_{1}^{(+)}(0,\x)\right)} \ .\]
Neglecting terms that vanish in the limit
$\z \to \x$, we obtain
\( 2L\frac{\partial}{\partial \x}
\left[\psi^{+}_{1}(0,\z)\psi_{1}(0,\x)\right] = \)
\bea
&&{\textstyle\left(\frac{1}{i\frac{\pi}{L}(\z-\x)^{2}}+{\rm const}. \right)}
e^{-\frac{i\pi}{2L}(\z-\x)}e^{-2\Lambda_{1}(\z-\x)}:e^{\left(\varphi_{1}
(0,\z)-\varphi_{1}(0,\x)\right)}: +\\ 
&&+{\textstyle\frac{1}{i\frac{\pi}{L}(\z-\x)}}
e^{-2\Lambda_{1}(\z-\x)}\Biggl\{
\left(-2{\textstyle\frac{\partial}
{\partial \x}}\Lambda_{1}(\z-\x)+{\textstyle\frac{i\pi}{2L}}
\right\}:e^{\left(\varphi_{1}(0,\z)-\varphi_{1}(0,\x)\right)}:-  \\
&&-\partial_{1}\varphi_{1}^{(-)}(0,\x)
:e^{\left(\varphi_{1}(0,\z)-\varphi_{1}(0,\x)\right)}:-
:e^{\left(\varphi_{1}(0,\z)-\varphi_{1}(0,\x)\right)}:
\partial_{1}\varphi_{1}^{(+)}(0,\x) \Biggr\}.
\eea 
By expanding in Taylor series one finds
\bean 
\label{perv1}
\psi^{+}_{1}(0,\x+\epsilon)\partial_{1}\psi_{1}(0,\x) &=&
\frac{i}{4\pi}:\Big(\partial_{1}\varphi_{1}(0,\x)\Big)^{2}:
-\frac{i\pi}{16L^{2}e^{2}}(Q+Q_{5})^{2} - \nn\\ 
&&-\frac{Q+Q_{5}}{4Le}\partial_{1}\varphi_{1}(0,\x)
-\frac{i}{4\pi}\partial_{1}^{2}
\varphi_{1}(0,\x)+\frac{1}{2i\pi\epsilon^{2}} +{\rm const}. \ .
\eean 
An analogous calculation gives
\bean 
\label{perv2}
\psi^{+}_{2}(0,\x+\epsilon)\partial_{1}\psi_{2}(0,\x) &=&
-\frac{i}{4\pi}:\Big(\partial_{1}\varphi_{2}(0,\x)\Big)^{2}:
+\frac{i\pi}{16L^{2}e^{2}}(Q-Q_{5})^{2} -
 \nn\\ 
&&-\frac{Q-Q_{5}}{4Le}\partial_{1}\varphi_{2}(0,\x)
+\frac{i}{4\pi}\partial_{1}^{2}
\varphi_{2}(0,\x)-\frac{1}{2i\pi\epsilon^{2}} +{\rm const}.\ . 
\eean 
In the previous formulae, positive and negative energy components
$\varphi_{i}^{(+)}$ \ and  \ $\varphi_{i}^{(-)}$ refer to the free
fermion theory. 

In the light-cone gauge the {\it left} component of the
fermionic field does not interact: in the full Hamiltonian $\varphi_1$
is decoupled. The coupling with the gauge field only affects
$\varphi_{2}=\varphi_{2}^{(+)}+
\varphi_{2}^{(-)}$: the two frequencies mix under time evolution.
The vacuum involved in the $v.e.v.$ subtraction refers to the
interacting theory.

If we define
\[H_{\psi}^{(0)}=\frac{1}{2}\int_{-L}^{L} d\x
:\Bigl[i\psi^{+}_{1}(0,\x)\partial_{1}\psi_{1}(0,\x)
-i\psi^{+}_{2}(0,\x)\partial_{1}\psi_{2}(0,\x)+h.c.\Bigr]: \ ,\]
where the dots denote normal ordering with respect to the above-mentioned
frequencies, we easily get
\[ H_{\psi}^{(0)}=
\frac{\pi}{4Le^{2}}(Q^{2}+Q_{5}^{2})
+\frac{\pi}{L}\som |n|\gamma_{n}^{+}\gamma_{n} \ .\]
$H_{\psi}^{(0)}$ takes care of the first two terms in eqs. (\ref{perv1})
and (\ref{perv2}). The third and fourth contributions vanish, thanks
to the periodicity of $\varphi$. The singular $1/\epsilon^2$ terms
require an expansion of the phase factor up to the second order:
\[
e^{-ie\int_{\x}^{\x+\epsilon}A_{1}(0,\y)d\y} =
1-ie A_{1}(0,\x)\epsilon-\frac{ie}{2}\partial_{1}A_{1}(0,\x)\epsilon^{2}
-\frac{e^{2}}{2}A_{1}^{2}(0,\x)\epsilon^{2}+\ldots \ .
\]
The term $A_{1}^{2}$ needs a suitable definition.

Putting everything together, one can realize that the single poles in $\epsilon$
cancel under Hermitian conjugation; double pole and finite constants, being 
{\it c} numbers, are subtracted by $v.e.v.$.

We eventually obtain
\[H_{\psi}= {\cal N}\left[\som k_{0}\gamma_{n}^{+}\gamma_{n}
-\int_{-L}^{L}d\x \frac{e^{2}}{2\pi}A_{1}^{2}(0,\x)\right]
+\frac{\pi}{4Le^{2}}(Q^{2}+Q_{5}^{2}) \ ,\]
or, using (\ref{A}), (\ref{F}):
\be \label{hpsi}
H_{\psi}= {\cal N}\left[\som k_{0}\gamma_{n}^{+}\gamma_{n}
- \frac{e^{2}}{2\pi}\som a_{n}a_{-n}\right]
+\frac{\pi}{4Le^{2}}(Q^{2}+Q_{5}^{2})-\frac{e^{2}}{2\pi}a_{0}^{2} \ .
\ee

The symbol ${\cal N}$ has been introduced because the products
appearing there need to be regularized.
Since the $v.e.v.$ involves the vacuum of the interacting theory,
we shall see in the next section that the ordering ${\cal N}$
is nothing but the normal ordering with respect to the operators
that diagonalize the full Hamiltonian. 

\section{Diagonalization of the interacting Hamiltonian}

We consider the classical Hamiltonian
(\ref{H}).
Using eqs. (\ref{j+}), (\ref{hpsi}), 
(\ref{A}) and  (\ref{F}), we get the following quantum expression:
\bean
\label{avarizia}
H&=&\frac{1}{2}{\bf b_{0}}^{2}+\frac{e^{2}}{2\pi}{\bf a_{0}}^{2}+
\frac{\pi}{4Le^{2}}(Q^{2}+Q_{5}^{2})+\frac{1}{\sqrt{2L}}{\bf a_{0}}(Q-Q_{5}) 
\\
&&+{\cal{N}}\Biggl[\frac{1}{2}\som b_{n}b_{-n}+
i\som k_{1}a_{n}b_{-n}
+\frac{e^{2}}{2\pi}\som a_{n}a_{-n}+ \nn \\
&&
+\frac{ie}{\sqrt{\pi}}\sneg \sqrt{2k_{0}}(\gamma_{n}a_{n}-\gamma_{n}^{+}
a_{-n})+\som k_{0}\gamma_{n}^{+}\gamma_{n}
\Biggr].\nn
\eean

Setting
$m=\frac{e}{\sqrt{\pi}}$, $ \omega=\sqrt{k_{0}^{2}+m^{2}} $ 
and defining, for $n<0$, the operators
\bean
\chi_{n}&=&\gamma_{n}^{+}-\frac{\sqrt{2\pi k_{0}}}{e}b_{n} \ , \nn\\
\Sigma_{n}&=&
\frac{1}{\sqrt{\omega}}\left(
\frac{\sqrt{\pi}}{e\sqrt{2}}(\omega+k_{0})b_{n}
-\frac{ie}{\sqrt{2\pi}}a_{n}
-\sqrt{k_{0}}\gamma_{n}^{+}
\right) \label{trasformazione} \ , \\
\Sigma_{-n}&=&
\frac{1}{\sqrt{\omega}}\left(
\frac{\sqrt{\pi}}{e\sqrt{2}}(\omega-k_{0})b_{-n}
-\frac{ie}{\sqrt{2\pi}}a_{-n}
+\sqrt{k_{0}}\gamma_{n}
\right) \ , \nn  
\eean 
it is easy to check the following commutation relations:
\be \label{piu`}
[\Sigma_{n},\Sigma_{m}^{+}]=\delta_{mn} \quad , \quad
[\Sigma_{-n},\Sigma_{-m}^{+}]=\delta_{mn} \ , 
\ee
\be \label{meno}
[\chi_{n},\chi_{m}^{+}]=-\delta_{mn} \ ,
\ee
all other commutators vanishing.
In terms of these operators, the frequency part of the Hamiltonian 
(\ref{avarizia}) takes the form

\bea
H_{1}=&&{\cal{N}}\Biggl[\frac{1}{2}\som b_{n}b_{-n}+
i\som k_{1}a_{n}b_{-n}
+\frac{e^{2}}{2\pi}\som a_{n}a_{-n}+\\ \\
&&\qquad
+\frac{ie}{\sqrt{\pi}}\sneg \sqrt{2k_{0}}(\gamma_{n}a_{n}-\gamma_{n}^{+}
a_{-n})+\som k_{0}\gamma_{n}^{+}\gamma_{n} \Biggr]= \\ \\
&&\qquad \quad 
=\spos k_{0}\gamma_{n}^{+}\gamma_{n}
-\sneg k_{0}\chi_{n}^{+}\chi_{n}
+\som \sqrt{k_{0}^{2}+m^{2}}\; \Sigma_{n}^{+}\Sigma_{n} \ .
\eea 
$H_{1}$ takes the form of the Fock Hamiltonian of a massive
boson field (apart from the zero mode) with
$m=\frac{e}{\sqrt{\pi}}$ and of two massless chiral bosons,
one of which having commutation relations leading to an
indefinite metric. We notice that the
term with negative sign in $H_{1}$ 
and the commutator of $\chi$ and $\chi^{+}$, entail that
$\gamma_{n}$, for $n>0$, and $\chi_{n}$ both evolve in time with
a positive frequency. As we shall see, this property allows us to express
the selection of a physical subspace in terms of annihilation operators.
Inserting in
(\ref{trasformazione}) the expressions of $a_{n}$ and $b_{n}$ 
in terms of the creation and annihilation operators of the gauge 
fields, we obtain, for $n<0$: 
\bea
\chi_{n}&=& \gamma_{n}^{+}+\frac{ik_{0}\sqrt{\pi}}{e}(A_{n}-A_{-n})\ , \\
\Sigma_{n}&=&-\sqrt{\frac{k_{0}}{\omega}}\gamma_{n}^{+}
-\frac{ie}{2\sqrt{\pi k_{0}\omega}}(A_{n}+A_{-n})
-\frac{i\sqrt{\pi k_{0}\omega}}{2e}\left(1+\frac{k_{0}}{\omega}\right)
(A_{n}-A_{-n})\ , \\
\Sigma_{-n}&=&\sqrt{\frac{k_{0}}{\omega}}\gamma_{n}
-\frac{ie}{2\sqrt{\pi k_{0}\omega}}(A_{n}^{+}+A_{-n}^{+})
-\frac{i\sqrt{\pi k_{0}\omega}}{2e}\left(1-\frac{k_{0}}{\omega}
\right)(A_{-n}^{+}-A_{n}^{+}) \ .
\eea 
One can verify that the following unitary operator:
\bea
U&=&e^{i \sum_{n<0} \frac{k_{0}\sqrt{\pi}}{e}\left[\gamma_{n}^{+}(A_{n}^{+}
-A_{-n}^{+})+\gamma_{n}(A_{n}-A_{-n})\right]} \\ 
&& \cdot e^{\sum_{n<0} \theta (A_{n}A_{-n}^{+}-A_{n}^{+}A_{-n}) }
e^{ \frac{\pi}{2}\sum_{n<0} (\gamma_{n}A_{-n}-A_{-n}^{+}\gamma_{n}^{+}) } \ ,
\eea
with
\[\theta=\ln\left(\frac{e}{\sqrt{\pi k_{0}\omega}} \right) \ ,\]
maps creation and annihilation operators of the free original
fields into new operators that diagonalize the interacting Hamiltonian.
As a matter of fact, for $n<0$, it can be checked that:
\bea
&&UA_{-n}U^{-1}=-\chi_{n} \ , \\ 
&&UA_{n}U^{-1}=i\Sigma_{n}  \ ,\\ 
&&U\gamma_{n}U^{-1}=i\Sigma_{-n} \ , \\ 
&&U\gamma_{-n}U^{-1}=i\gamma_{-n} \ .
\eea 
It is easy to realize that from
\[A_{p}|M,N\rangle=\gamma_{p}|M,N\rangle=0 \qquad , \qquad p\neq 0 \ ,\] 
it follows that
\be \label{mn}
\Sigma_{\pm p}U|M,N\rangle=\chi_{p}U|M,N\rangle=
\gamma_{-p}U|M,N\rangle=0 \qquad , \qquad
p<0 \ .
\ee
Therefore $U$ maps the base vectors of the free-field Fock space
obtained by repeated action of
$A^{+}_{p}$ and $\gamma^{+}_{p}$ on the states $|M,N\rangle$, in
eigenstates of $H_{1}$ obtained by repeated action of $\Sigma^{+}_{\pm p}$,
$\chi^{+}_{p}$ and $\gamma_{-p}^{+}$ $(p<0)$ on the states $U|M,N\rangle$.

One obtains the zero mode of the massive boson by diagonalizing
\[ H_{0}=\frac{1}{2}{\bf b_{0}}^{2}+\frac{e^{2}}{2\pi}{\bf a_{0}}^{2}+
\frac{\pi}{4Le^{2}}(Q^{2}+Q_{5}^{2})
+\frac{1}{\sqrt{2L}}{\bf a_{0}}(Q-Q_{5}) \ .\]
By defining
\bean
\label{modinulli}
&&\Sigma_{0}=\alpha
+\frac{Q-Q_{5}}{2i\sqrt{L}m^{\frac{3}{2}}}=
\frac{1}{\sqrt{2}}\left(
-i\sqrt{m}{\bf a_{0}}+\frac{\bf {b_{0}}}{\sqrt{m}}
+\frac{Q-Q_{5}}{i\sqrt{2L}m^{\frac{3}{2}}}\right)\ , \nn\\ 
&&\Sigma_{0}^{+}=\alpha^{+}
-\frac{Q-Q_{5}}{2i\sqrt{L}m^{\frac{3}{2}}}=
\frac{1}{\sqrt{2}}\left(
i\sqrt{m}{\bf a_{0}}+\frac{{\bf b_{0}}}{\sqrt{m}}
-\frac{Q-Q_{5}}{i\sqrt{2L}m^{\frac{3}{2}}}\right) \ ,
\eean 
the commutation relations 
$[\alpha,\alpha^{+}]=1$, $[\Sigma_{0}, \Sigma_{0}^{+}]=1$ 
follow and the Hamiltonian $H_0$ can be written as
\[H_{0}=\frac{e}{\sqrt{\pi}}\Sigma_{0}^{+}\Sigma_{0}
+\frac{\pi}{2e^{2} L}QQ_{5} \ .\]
We are thus led to the following diagonal form of the complete
Hamiltonian:
\[H=\spos k_{0}\gamma_{n}^{+}\gamma_{n}
 -\sneg k_{0}\chi_{n}^{+}\chi_{n}
+\somma \sqrt{k_{0}^{2}+m^{2}}\; \Sigma_{n}^{+}\Sigma_{n}
+\frac{\pi}{2e^{2} L}QQ_{5} \ ,\] 
$\alpha$ and $\Sigma_{0}$ being related by a unitary operator:
\[\Sigma_{0}=U_{0}\alpha U_{0}^{-1} \ ,\]
where 
\[ U_{0}=
e^{\frac{i}{2\sqrt{L}m^{3/2}}(Q-Q_{5})(\alpha+\alpha^{+})} \ .\]
We remark that $Q$ , $Q_{5}$ and the spurion
$\sigma_{1}$ commute with $U_{0}$, whereas
\be \label{sigma2}
\underline{\sigma}_{2}=U_{0}\sigma_{2}U_{0}^{-1}=
e^{-ie\frac{1}{\sqrt{L}m^{3/2}}(\alpha+\alpha^{+})}\sigma_{2}.
\ee
We have the following commutators between the operators $\sigma_{1}$ , 
$\underline{\sigma}_{2}$ , $Q$ , $Q_{5}$ , $\Sigma_{0}$
$\Sigma_{0}^{+}$ 
\[
[\Sigma_{0}, \Sigma_{0}^{+}]=1 \ ,
\]
\[
\{\sigma_{1},\underline{\sigma}_{2}\}=0 \ ,
\]
\bean
&& [Q,\sigma_{1}]=-e\sigma_{1}\qquad, \qquad [Q_{5},\sigma_{1}]=-e\sigma_{1} \ ,
\nn \\ 
&&
[Q,\underline{\sigma}_{2}]=-e\underline{\sigma}_{2} \qquad, \qquad
[Q_{5},\underline{\sigma}_{2}]=e\underline{\sigma}_{2} \ ,
\label{commutatori}
\eean 
all other commutators vanishing.

Starting from the classical expression of the momentum
\bea
P\equiv P^{1}=\int_{-L}^{L}d\x \Theta^{01}
&=&\int_{-L}^{L}d\x\Biggl[\frac{i}{2}\Bigl(\psi^{+}(0,\x)\partial^{1}\psi(0,\x)
-\partial^{1}\psi^{+}(0,\x)\psi(0,\x)\Bigr)+
\\ 
&+&\frac{1}{\sqrt{2}}F(0,\x)\partial^{1}A(0,\x)\Biggr] \ ,
\eea
the same transformation (\ref{trasformazione}) leads to the 
following diagonal expression 
\[P=-\spos k_{0} \gamma_{n}^{+}\gamma_{n}
+\sneg k_{0}\chi_{n}^{+}\chi_{n}+
\som k_{1}\Sigma_{n}^{+}\Sigma_{n}
-\frac{\pi}{2Le^{2}}QQ_{5} \ .\] 
We notice that the modes $\gamma$ and $ \chi$ have a negative momentum
(owing to the negative sign in the $\chi$ commutator).
Therefore they describe bosonic degrees of freedom of the same {\it left}
chirality.

\section{The temporal evolution}

The temporal evolution can easily be determined, starting from the commutation
relations
\bean \label{cupidigia}
&&[H,\Sigma_{0,\pm n}]=-\omega \Sigma_{0,\pm n}
\qquad , \qquad 
[H,\Sigma^{+}_{0,\pm n}]=\omega \Sigma^{+}_{0,\pm n} \ , \nn \\
&&[H,\chi_{n}]=-k_{0}\chi_{n} \qquad \quad  \; ,\;\quad \qquad
[H,\chi_{n}^{+}]=k_{0}\chi_{n}^{+} \ ,\nn  \\ 
&&[H,\gamma_{-n}]=-k_{0}\gamma_{-n} \qquad  \;\; ,\;\; \qquad
[H,\gamma^{+}_{-n}]=k_{0}\gamma^{+}_{-n} \ , \nn \\ \\
&&[H,\sigma_{1}]=-\frac{\pi}{4eL}\Bigl[(Q+Q_{5})\sigma_{1}
+\sigma_{1}(Q+Q_{5})\Bigr] \ , \nn\\ 
&&[H,\underline{\sigma}_{2}]=\frac{\pi}{4eL}\Bigl[(Q-Q_{5})
\underline{\sigma}_{2}+
\underline{\sigma}_{2}(Q-Q_{5})\Bigr] \ ,\nn \\ 
&&[H,Q]=[H,Q_{5}]=0 \ .\nn
\eean 
Setting
\be \label{Sigma(x)} 
\Sigma(t,\x)=\frac{1}{\sqrt{2L}}\somma
\frac{1}{\sqrt{2\omega}} \left(
\Sigma_{n}e^{-i(\omega t- k_{1}\x)}+\Sigma_{n}^{+}
e^{i(\omega t - k_{1}\x)}\right) 
\ee
and
\be \label{puledra}
\chi(t,\x)=\frac{1}{\sqrt{2L}}\sneg \frac{1}{\sqrt{2k_{0}}} 
\left(\chi_{n}e^{-ik_{0}(t+\x)}+\chi_{n}^{+}e^{ik_{0}(t+\x)}\right),
\ee
from (\ref{cupidigia}) and the canonical commutation relations, it is easy 
to see that  
$\Sigma(t,\x)$ satisfies the 
Klein--Gordon equation:
\be \label{Klein-Gordon}
(\Box +m^{2})\Sigma(t,\x)=0
\ee
and
\[\Bigl[\Sigma(t,\x),\partial_{0}\Sigma(t,\y)\Bigr]=i\delta(\x-\y)
\quad , \quad \Bigl[\Sigma(t,\x),\Sigma(t,\y)\Bigr]=
\Bigl[\partial_{0}\Sigma(t,\x),\partial_{0}\Sigma(t,\y)\Bigr]=0 \ .\]
Here $\chi(t,\x)$ is a massless chiral (left) field
satisfying the algebra
\[\Bigl[{\chi}(t,\x),\partial_{0}{\chi}(t,\y)\Bigr]
={-i\over {2L}} \sneg e^{-\epsilon k_0} \cos k_0 (\x - \y) \ ,\]
\[\Bigl[{\chi}(t,\x),{\chi}(t,\y)\Bigr]={i\over {2L}}\sneg {{\sin k_0 (\x - \y)}
\over {k_0}} \ ,\]
\[\Bigl[\partial_{0}{\chi}(t,\x),\partial_{0}{\chi}(t,\y)\Bigr]= {i\over{2L}}
\sneg k_0 \sin k_0 (\x - \y) \ .\]
The temporal evolution of the original fields can now be obtained in the 
Heisenberg picture in the usual way; for instance
\[A(t,\x)=e^{iHt}A(0,\x)e^{-iHt} \ ,\]
and analogous ones for the other fields.
We easily get
\be \label{A(x)}
A(t,\x)=-\frac{2}{m}\partial_{+}\Bigl[\Sigma(t,\x)+\chi(t,\x)\Bigr]
-\frac{Q-Q_{5}}{\sqrt{2}m^{2}L},
\ee
which in turn implies
\be \label{Sigma-F}
F(t,\x)=m\Sigma(t,\x).
\ee
Therefore the electric field $F$ becomes massive
\[(\Box +m^{2})F(t,\x)=0.\]
We now turn our attention to the fermionic field. As far as the $\psi_{1}$
component is concerned, one immediately recovers a free temporal evolution: 
\be \label{psi1}
 \psi_{1}(t,\x)=\textstyle{\frac{1}{\sqrt{2L}}}e^{-\varphi_{1}^{(-)}(t,\x)}
e^{-\frac{i\pi}{4eL}(Q+Q_{5})(t+\x)}\sigma_{1}
e^{-\frac{i\pi}{4eL}(Q+Q_{5})(t+\x)}
e^{-\varphi_{1}^{(+)}(t,\x)} \ ,
\ee
\bea
&&\varphi_{1}^{(-)}(t,\x)=-\spos \frac{i}{\sqrt{n}}\gamma_{n}^{+}
e^{ik_{0}(t+\x)} \ , \\ 
&&\varphi_{1}^{(+)}(t,\x)=-\spos \frac{i}{\sqrt{n}}\gamma_{n}
e^{-ik_{0}(t+\x)} \ .
\eea
For the $\psi_{2}$ component we get instead
\[\psi_{2}(0,\x)=\frac{1}{\sqrt{2L}}
\exp\left\{i\sneg \frac{1}{\sqrt{|n|}}\left[\chi_{n}
+\sqrt{\frac{k_{0}}{\omega}}\left(\Sigma_{n}+\Sigma_{-n}^{+}\right)\right]
e^{ik_{1}\x}  \right\}\cdot \]
\[\quad \qquad \cdot e^{i\frac{\pi}{4Le}(Q-Q_{5})\x}
e^{i\sqrt{\frac{\pi}{mL}}\Sigma_{0}^{+}}
\underline{\sigma}_{2} 
e^{i\sqrt{\frac{\pi}{mL}}\Sigma_{0}}
e^{i\frac{\pi}{4Le}(Q-Q_{5})\x}e^{-\frac{\pi}{2mL}}\cdot \]
\[\; \quad
\cdot \exp\left\{i\sneg \frac{1}{\sqrt{|n|}}\left[\chi_{n}^{+}
+\sqrt{\frac{k_{0}}{\omega}}(\Sigma^{+}_{n}+\Sigma_{-n})\right]
e^{-ik_{1}\x}  \right\}=\]
\[\qquad \qquad \qquad
=\frac{e^{Z}}{\sqrt{2L}}
\exp\left\{i\sneg \frac{1}{\sqrt{|n|}}\left[\chi_{n}^{+}e^{-ik_{1}\x}
+\sqrt{\frac{k_{0}}{\omega}}\left(\Sigma_{n}^{+}e^{-ik_{1}\x}
+\Sigma_{-n}^{+}e^{ik_{1}\x}\right)  \right]
\right\} \cdot\]
\[ \cdot e^{i\frac{\pi}{4Le}(Q-Q_{5})\x}
e^{i\sqrt{\frac{\pi}{mL}}\Sigma_{0}^{+}}
\underline{\sigma}_{2} 
e^{i\sqrt{\frac{\pi}{mL}}\Sigma_{0}}
e^{i\frac{\pi}{4Le}(Q-Q_{5})\x}\cdot\]
\[ \quad \;
\cdot \exp\left\{i\sneg \frac{1}{\sqrt{|n|}}\left[\chi_{n}e^{ik_{1}\x}
+\sqrt{\frac{k_{0}}{\omega}}\left(\Sigma_{n}e^{ik_{1}\x}
+\Sigma_{-n}e^{-ik_{1}\x}\right)  \right] \right\} \ ,\]
where
\bean
Z&\!=&\!-\frac{\pi}{2mL}-\Biggl[\; \sneg \frac{1}{\sqrt{|n|}}
\chi_{n}e^{in\frac{\pi}{L}\x}\;\; ,\;\;
\sum_{p=-\infty}^{-1}\frac{1}{\sqrt{|p|}}
\chi_{p}^{+}e^{-ip\frac{\pi}{L}\x}\;\Biggr] + \nn \\ 
&&\!+\frac{\pi}{L}\Biggl[\;\sneg \frac{1}{\sqrt{\omega}}
\Sigma_{n}e^{in\frac{\pi}{L}\x} \;\;,\;\; 
\sum_{p=-\infty}^{-1}\frac{1}{\sqrt{\omega}}
\Sigma_{p}^{+}e^{-ip\frac{\pi}{L}\x}\;\Biggr] =
\nn \\ 
&\!=&\!-\frac{\pi}{2mL}
+\!\sneg\left( \frac{1}{|n|}-\frac{\pi}{L}\frac{1}{\omega}\right)
=\,-\frac{\pi}{2mL}+\spos\frac{m^{2}}{n\sqrt{\left(\frac{n\pi}{L}\right)^{2}
+m^{2}}\left(\sqrt{\left(\frac{n\pi}{L}\right)^{2}
+m^{2}}+n\frac{\pi}{L}  \right) }
\eean
is a finite constant.\footnote{We remark that, in the decompactification limit
$L \to \infty$, the factor ${{e^Z}\over {\sqrt {2L}}}$ would diverge, producing
an infinite renormalization constant for the field $\psi_2$.} 

Using the relation
\[e^{iHt}\underline{\sigma}_{2}e^{-iHt}=
e^{\frac{i\pi}{4Le}(Q-Q_{5})t} \underline{\sigma}_{2}
e^{\frac{i\pi}{4Le}(Q-Q_{5})t} \ ,\]
one can eventually find the temporal evolution of
$\psi_{2}$ :
\( \psi_{2}(t,\x)=e^{iHt}\psi_{2}(0,\x)e^{-iHt}= \)
\bean
&=&\frac{e^{Z}}{\sqrt{2L}}
e^{2i\sqrt{\pi}\left[\chi^{(-)}(t,\x)+\Sigma^{(-)}(t,\x) \right] }
e^{\frac{i\pi}{4Le}(Q-Q_{5})(t+\x)} \;\underline{\sigma}_{2}\cdot \nn \\
&&\cdot e^{\frac{i\pi}{4Le}(Q-Q_{5})(t+\x)}
e^{2i\sqrt{\pi}\left[\chi^{(+)}(t,\x)+\Sigma^{(+)}(t,\x)\right] } \ ,
\label{psi2}
\eean
where
\bea
&&\chi^{(-)}(t,\x)=\frac{1}{\sqrt{2L}}\sneg \frac{1}{\sqrt{2k_{0}}} 
\chi_{n}^{+}e^{ik_{0}(t+\x)}\ ,\\ 
&&\chi^{(+)}(t,\x)=\frac{1}{\sqrt{2L}}\sneg \frac{1}{\sqrt{2k_{0}}} 
\chi_{n}e^{-ik_{0}(t+\x)}\ , \\ 
&&\Sigma^{(-)}(t,\x)=\frac{1}{\sqrt{2L}}\somma
\frac{1}{\sqrt{2\omega}}\Sigma_{n}^{+}
e^{i(\omega t - k_{1}\x)}\ ,\\ 
&&\Sigma^{(+)}(t,\x)=\frac{1}{\sqrt{2L}}\somma
\frac{1}{\sqrt{2\omega}} 
\Sigma_{n}e^{-i(\omega t- k_{1}\x)} \ .
\eea 
It can be checked that the field 
$\psi_{2}$ satisfies the regularized
Dirac equation (\ref{checca}).

Finally, for 
$\lambda$  we get
\bea
\lambda(t,\x)&=&e^{iHt}\lambda(0,\x)e^{-iHt} =\\ 
&=&\frac{Q}{\sqrt{2}L}+\frac{m}{\sqrt{2L}}
\sneg \sqrt{k_{0}}\left[i\left(\gamma_{-n}-\chi_{n}\right)
e^{-ik_{0}(t+\x)}-i\left(\gamma_{-n}^{+}-\chi_{n}^{+} \right)
e^{ik_{0}(t+\x)}  \right] \ ,
\eea 
in agreement with both Euler--Lagrange equations for the electric field;
$\lambda$ cannot vanish strongly as it has non-vanishing
commutators with $A$ and $\psi$.
In order to recover the correct Maxwell equations in the weak sense 
we have to define the 
physical states by imposing the following conditions
\bean
&&Q|phys\rangle=0 \ , \\ \label{Q}
&&\lambda^{(+)}|phys\rangle =0 \ ,\label{lambda+}
\eean
$\lambda^{(+)}$ being the positive frequency component of
$\lambda(x)$.

Equation (\ref{lambda+}) is equivalent to
\be \label{phys2}
(\gamma_{-n}-\chi_{n})|phys\rangle=0 \quad ,\quad \forall n<0.
\ee

\section{The structure of the Hilbert space}
\subsection{The vacuum sector}

The  operators
$H$, $P$, $Q$ and $Q_{5}$ commute pair-wise. The common eigenstates
can be obtained by the repeated action of the creation operators
$\Sigma_{0,\pm p}^{+}$ , $\gamma_{-p}^{+}$ , $\chi^{+}_{p} $
($p<0$) to the states
\[ |\Omega_{mn}\rangle =S\sigma_{1}^{m}\sigma_{2}^{n}
|0\rangle = \sigma_{1}^{m}\underline{\sigma}_{2}^{n}S
|0\rangle\ , \qquad m,n=0,\pm 1,\pm 2, \ldots \ \ ,\]
where $|0\rangle$ is defined by the conditions
\[ \alpha|0\rangle=A_{n}|0\rangle=\gamma_{n}|0\rangle
=Q|0\rangle=Q_{5}|0\rangle=0\]
and
\[S=UU_{0}\ .\]
$|\Omega_{mn}\rangle$ obeys the equations
\[\Sigma_{0,\pm p}|\Omega_{mn}\rangle =
\gamma_{-p}|\Omega_{mn}\rangle =\chi_{p}|\Omega_{mn}\rangle  =0 \ ,\]
\bea
\langle \Omega_{mn}|\Omega_{pq}\rangle&=&
\langle 
0|\sigma_{2}^{-n}\sigma_{1}^{-m}S^{+}S\sigma_{1}^{p}\sigma_{2}^{q}
|0\rangle=(-1)^{(p-m)n} \langle 0|\sigma_{1}^{(p-m)}\sigma_{2}^{(q-n)}
|0\rangle=\\ 
&&=(-1)^{(p-m)n} \langle 0|p-m,q-n\rangle= \delta_{pm}
\delta_{qn }\langle 0|0\rangle \ ,
\eea 
so that
\[\langle \Omega_{mn}|\Omega_{pq}\rangle=
\delta_{pm}\delta_{qn }\ , \quad {\rm if} \quad 
\langle 0|0\rangle=1 \ ,\]
and
\bea
&&Q|\Omega_{mn}\rangle=-e(m+n)|\Omega_{mn}\rangle \ , \\
&&Q_{5}|\Omega_{mn}\rangle =-e(m-n)|\Omega_{mn}\rangle \ , \\
&&H|\Omega_{mn}\rangle =\frac{\pi}{2L}(m^{2}-n^{2})|\Omega_{mn}\rangle \ , \\
&&P|\Omega_{mn}\rangle =-\frac{\pi}{2L}(m^{2}-n^{2})|\Omega_{mn}\rangle \ .
\eea
Among them, only the states
\[|\Omega_{-n\,n}\rangle=S\sigma_{1}^{-n}\sigma_{2}^{n}|0\rangle\]
satisfy the condition
\[ Q|\Omega_{-n\,n}\rangle =0 \ ,\]
and therefore are physical.

For such states
\[ Q_{5}|\Omega_{-n \, n}\rangle=2n|\Omega_{-n \, n }\rangle \ ,\]
\[ H|\Omega_{-n\,n}\rangle 
=P|\Omega_{-n\,n}\rangle =0 \ .\]
An infinite set of vacua, labelled by the integer $n$, are 
translation-invariant. 
Such states are related by a residual gauge transformation.
As a matter of fact, if we define
\[ |\Omega_{n}\rangle=e^{i\pi\frac{n(n-1)}{2}} 
|\Omega_{-n\,n}\rangle =
e^{i\pi\frac{n(n-1)}{2}} \sigma_{1}^{-n}\underline{\sigma}_{2}^{n}
S|0\rangle =(\sigma_{1}^{-1}\underline{\sigma}_{2})^{n}S|0\rangle \ ,\]
we get
\[ T_{1}|\Omega_{n}\rangle=|\Omega_{n+1}\rangle \ ,\]
with
\[ T_{1}=\sigma_{1}^{-1}\underline{\sigma}_{2} \ .\]
The unitary operators $T_{m}=(T_{1})^{m}$ generate residual gauge 
transformations with
winding number $m$ :
\[ T_{m}A(x)T_{m}^{+}=A(x)-\frac{m\pi\sqrt{2}}{eL} \ ,\]
\[ T_{m}\psi_{\alpha}(x)T_{m}^{+}=(-1)^{m}e^{im\frac{\pi}{L}(t+\x)}
\psi_{\alpha}(x) \ .\]
In order to construct gauge-invariant vacua (up to a phase factor),
$|\theta\rangle$ vacua are introduced: 
\[ |\theta\rangle=\somma e^{in\theta}|\Omega_{n}\rangle \ ,\]
which diagonalize the $T_{m}$ operators
\[ T_{m}|\theta\rangle=e^{-im\theta}|\theta\rangle \ .\]
$|\theta\rangle$ vacua are necessary to recover the validity
of cluster decomposition. Different values of $\theta$ give
rise to inequivalent representations. 

\subsection{The physical subspace}
We notice that the indefinite inner product
$(\cdot ,\cdot)$, which is preserved by the $U$ operator,
entails the presence of negative norm states.
On the other hand, the scalar product $\langle \cdot , \cdot \rangle$,
is not preserved by $U$
and the relations (\ref{piu`}), (\ref {meno}) no longer hold for the
scalar product $\langle \cdot,\cdot \rangle$, if the adjoint operation
is defined according to the indefinite inner product.

A Fock space ${\cal V}$ can be constructed by repeated action
of creation operators
$\Sigma^{+}_{\pm,0}$, $\chi^{+}_{n}$
$\gamma^{+}_{-n}$ $(n<0)$ on a $|\theta\rangle$ vacuum, endowed with
a positive-definite inner product
\ $(\cdot ,\cdot )_{+}$, which preserves eqs.
(\ref{piu`}) and such that the commutator (\ref{meno}) becomes canonical
if $\chi^{+}$ is replaced by the adjoint of
$\chi$ with respect to $(\cdot , \cdot )_{+}$ . In this case
\[ (\cdot,\cdot)=(\cdot,\eta'\cdot)_{+}
\qquad , \quad {\rm with} \quad \eta'=(-1)^{N_{\chi}} \ ,\]
where
\[ N_{\chi}=-\sneg \chi_{n}^{+}\chi_{n} \ . \]

If we select the physical space ${\cal V}_{p}$ with the condition
(\ref{phys2}), we can easily see that the eigenstates $|\phi_{n}\rangle$
of the operator 
\[ N_{\chi\gamma} =\sneg \left(\gamma^{+}_{-n}\gamma_{-n}
-\chi^{+}_{n}\chi_{n}\right) \]
\[ N_{\chi\gamma}|\phi_{n}\rangle=n|\phi_{n}\rangle \ ,\]
have vanishing norm for $n\neq 0$ and do not contribute to physical quantities.
The physical Hilbert space can be defined as usual as a quotient space.

\subsection{Chiral anomaly and fermionic condensate}

From the temporal evolution of the fermion fields, it is easy to derive
the expressions for the currents
\bea
j^{0}(x)=j^{1}_{5}(x)&=&\frac{Q}{2L}+\frac{ie}{2\sqrt{\pi L}}
\som \frac{k_{1}}{\sqrt{\omega}}\left(\Sigma_{n}
e^{-i(\omega t-k_{1}\x)}-\Sigma_{n}^{+}e^{i(\omega t-k_{1}\x)}
\right) +\\ 
&&+\frac{ie}{2\sqrt{\pi L}}\sneg \sqrt{k_{0}}
\left[(\gamma_{-n}-\chi_{n})e^{-ik_{0}(t+\x)}-
(\gamma_{-n}^{+}-\chi_{n}^{+})e^{ik_{0}(t+\x)} \right] \ ,
\eea 
\bea
j^{1}(x)=j^{0}_{5}(x)&=&-\frac{Q_5}{2L}+\frac{ie}{2\sqrt{\pi L}}
\somma \sqrt{\omega}\left(\Sigma_{n}
e^{-i(\omega t-k_{1}\x)}-\Sigma_{n}^{+}e^{i(\omega t-k_{1}\x)}
\right) - \\
&&-\frac{ie}{2\sqrt{\pi L}}\sneg \sqrt{k_{0}}
\left[(\gamma_{-n}-\chi_{n})e^{-ik_{0}(t+\x)}-
(\gamma_{-n}^{+}-\chi_{n}^{+})e^{ik_{0}(t+\x)} \right] \; .
\eea 
Using (\ref{Sigma(x)}), 
(\ref{Klein-Gordon}) and (\ref{Sigma-F}), we get
\[ \partial_{\mu}j^{\mu}_{5}(x) =\frac{e^{2}}{\pi}F(x) \ .\]
As a consequence the chiral charge
\[ q_{5}=-\int_{-L}^{L}d\x j^{0}_{5}(x)= Q_5-im^{\frac{3}{2}}\sqrt{L}
\left( \Sigma_{0}e^{-imt}-\Sigma_{0}^{+} e^{imt} \right)\]
is not conserved and therefore cannot be the
generator of  chiral transformations. Instead, the generator is 
the ``free'' chiral charge $Q_{5}$, which commutes with the Hamiltonian.
On the other hand $q_5$ is gauge-invariant, while $Q_5$ is not.
The $|\theta\rangle$ vacua  are not invariant under the action of $Q_5$,
leading to a spontaneous chiral symmetry breaking
\[ e^{\frac{i}{e}\alpha Q_{5}}|\theta\rangle=
|\theta+2\alpha\rangle \; .\]
Let us now introduce the fermionic condensate
\[ \frac{\langle\theta|\bar{\psi}(x)\psi(x)|\theta\rangle}
{\langle\theta|\theta\rangle}
=\frac{1}{\langle\theta|\theta\rangle}
\Bigl[\langle\theta|\psi_{1}^{+}(x)\psi_{2}(x)|\theta\rangle
+\langle\theta|\psi_{2}^{+}(x)\psi_{1}(x)|\theta\rangle \Bigr] \ .\]
Using (\ref{psi1}) and (\ref{psi2}) it is easy to see that
\bea
\langle\theta|\psi_{1}^{+}(x)\psi_{2}(x)|\theta\rangle &=&
\frac{e^{Z}}{2L}
\langle\theta|e^{\frac{i\pi}{2Le}Q(t+\x)}
\sigma_{1}^{+}\underline{\sigma}_{2}
e^{\frac{i\pi}{2Le}Q(t+\x)}|\theta\rangle =\\ 
&&=\frac{e^{Z}}{2L}
\langle\theta|\sigma_{1}^{+}\underline{\sigma}_{2}
|\theta\rangle =\frac{e^{Z}}{2L}e^{-i\theta}{\langle\theta|\theta\rangle} \ ,
\eea
\bea
\langle\theta|\psi_{2}^{+}(x)\psi_{1}(x)|\theta\rangle &=&
\frac{e^{Z}}{2L}
\langle\theta|e^{-\frac{i\pi}{2Le}Q(t+\x)}
\underline{\sigma}_{2}^{+}\sigma_{1}
e^{-\frac{i\pi}{2Le}Q(t+\x)}|\theta\rangle =\\ 
&&=\frac{e^{Z}}{2L}
\langle\theta|\underline{\sigma}_{2}^{+}\sigma_{1}
|\theta\rangle =\frac{e^{Z}}{2L}e^{i\theta}{\langle\theta|\theta\rangle}
\eea
and, therefore,
\[ \frac{\langle\theta|\bar{\psi}(x)\psi(x)|\theta\rangle}
{\langle\theta|\theta\rangle}
=\frac{e^{Z}}{L}\cos \theta \ .\]
In the decompactification limit $L\rightarrow \infty$
one recovers the well-known result
\cite{MCC96}:
\[ \lim_{L\to \infty}\frac{e^{Z}}{L}\cos \theta=
\frac{m}{2\pi}e^{\gamma}\cos \theta \ ,\]
$\gamma$ being the Euler--Mascheroni constant.

\vskip 1.0truecm

\acknowledgments
We thank Gary McCartor for many useful discussions.
 
\def\ap {{\it Ann.\ Phys.\ (N.Y.)} }
\def\cmp {{\it Commun.\ Math.\ Phys.} }
\def\ijmpA {{\it Int.\ J.\ Mod.\ Phys.} { \bf A}}
\def\ijmpB {{\it Int.\ J.\ Mod.\ Phys.} { \bf B}}
\def\jmp {{\it  J.\ Math.\ Phys.} }
\def\mplA {{\it Mod.\ Phys.\ Lett.} { \bf A}}
\def\mplB {{\it Mod.\ Phys.\ Lett.} { \bf B}}
\def\plB {{\it Phys.\ Lett.} { \bf B}}
\def\plA {{\it Phys.\ Lett.} { \bf A}}
\def\nc {{\it Nuovo Cimento} }
\def\npB {{\it Nucl.\ Phys.} { \bf B}}
\def\pr {{\it Phys.\ Rev.} }
\def\prl {{\it Phys.\ Rev.\ Lett.} }
\def\prB {{\it Phys.\ Rev.} { \bf B}}
\def\prD {{\it Phys.\ Rev.} { \bf D}}
\def\prp {{\it Phys.\ Report} }
\def\ptp {{\it Prog.\ Theor.\ Phys.} }
\def\rmp {{\it Rev.\ Mod.\ Phys.} }
\def\hep {{\tt hep-th/}}


\begin{thebibliography}{99}

\bibitem{schwinger}
J. Schwinger, \pr {\bf 128}, 2425  (1962).
\bibitem{lowenstein}
J.H. Lowenstein and J.A. Swieca, \ap {\bf 68}, 172  (1971).
\bibitem{bas4}
A. Bassetto, M. Dalbosco, I. Lazzizzera and R. Soldati,
\prD {\bf 31}, 2012 (1985).
\bibitem{bas1}
A. Bassetto, G. Nardelli and R. Soldati, {\it Yang-Mills theories in algebraic
non-covariant gauges} (World Scientific, Singapore, 1991).
\bibitem{nakawaki}
Y. Nakawaki,
\ptp {\bf 64}, 1828 (1980); 
Y. Nakawaki,
\ptp {\bf 70}, 1105 (1983).
\bibitem{uhlenbrock}
D.A. Uhlenbrock,
\cmp {\bf 4}, 64 (1967). 
\bibitem{johnson}
K. Johnson,
{\it Nucl. Phys.} {\bf 25}, 431 (1961).
\bibitem{MCC96}
G. McCartor, \ijmpA{\bf 12}, 1091 (1997)
and references therein.

\end{thebibliography}
\end{document}